\documentclass[12pt]{article}

\usepackage[dvips]{graphicx}
\usepackage{epsfig}
\usepackage{amsmath,amsfonts,amssymb,amsthm}
\usepackage{verbatim}
\usepackage{psfrag}
\usepackage{bm}
\usepackage{bbm}
\usepackage[square,comma,sort&compress,numbers]{natbib}
\usepackage{color}
\usepackage{slashed}
\usepackage{color}

\usepackage{epsf,epsfig}
\usepackage{graphics}

\newcommand{\beq}{\begin{equation}}
\newcommand{\eeq}{\end{equation}}
\newcommand{\bea}{\begin{eqnarray}}
\newcommand{\eea}{\end{eqnarray}}

\newcommand{\bmatr}{\begin{pmatrix}}
\newcommand{\ematr}{\end{pmatrix}}

\newcommand{\la}{\langle}
\newcommand{\ra}{\rangle}

\newcommand{\gsim}
{\;\raisebox{-.3em}{$\stackrel{\displaystyle >}{\sim}$}\;}

\begin{document}
\thispagestyle{empty}

\begin{flushright}
{
\small
TUM-HEP-971-14\\
}
\end{flushright}

\vspace{0.4cm}
\begin{center}
\Large\bf\boldmath
Feynman Diagrams 
for Stochastic Inflation and
\\Quantum Field Theory in de Sitter Space
\unboldmath
\end{center}

\vspace{0.4cm}

\begin{center}
{Bj\"orn~Garbrecht$^a$, Florian Gautier$^a$, Gerasimos Rigopoulos$^{b}$ and Yi Zhu$^a$}\\
\vskip0.3cm
{\it $^a$Physik Department T70, James-Franck-Stra{\ss}e,\\
Technische Universit\"at M\"unchen, 85748 Garching, Germany\\
\vskip0.1cm
$^b$Institut f\"ur Theoretische Physik, Philosophenweg 12,\\
Universit\"at Heidelberg, 69120 Heidelberg, Germany}\\
\vskip.3cm
\end{center}

\begin{abstract}
\noindent We consider a massive scalar field with quartic self-interaction
$\lambda/4!\,\phi^4$ in de~Sitter spacetime and present a diagrammatic
expansion that describes the field as driven by stochastic noise. This is compared
with the Feynman diagrams in the Keldysh basis of the Amphichronous (Closed-Time-Path) Field Theoretical
formalism.
 For all orders
in the expansion, we find that the diagrams agree when evaluated
in the leading infrared approximation,
{\it i.e.} to leading order in $m^2/H^2$, where $m$ is the mass of the scalar field
and $H$ is the Hubble rate. As a consequence, the correlation
functions computed in both approaches also agree to leading infrared order.
This perturbative correspondence shows that the stochastic Theory is exactly equivalent to the Field Theory in the infrared. The former can then offer a non-perturbative resummation of the Field Theoretical Feynman diagram expansion, including fields with $0\leq m^2\ll\sqrt \lambda H^2$ for which the perturbation expansion fails at late times.
\end{abstract}

\section{Introduction}

The stochastic approach to Inflation~\cite{Starobinsky:1986fx,Starobinsky:1994bd}
is a simple and effective framework that can be used in order to
evaluate correlation functions of scalar fields in de~Sitter space on scales exceeding the horizon.
It can be derived from the underlying Field Theoretical formulation,  by treating
the short-wavelength modes as quantum noise to the horizon-size
field which is described as a classical random variable. This is justified by the
fact that the canonical commutator (between the field and the canonical momentum) estimated within the stochastic framework is small compared to the anti-commutator,
{\it i.e.} by the usual criterion for the classical behaviour of a dynamic system.
The resulting random walk of the scalar field (on top of the solution to the deterministic
equation of motion) does not only offer valuable intuition for understanding the field
evolution and the emergence of classical stochastic perturbations in the Universe, it is also useful in order to derive quantitative results~\cite{Martin:2005ir, Prokopec:2007ak, Riotto:2008mv, Enqvist:2011pt, Martin:2011ib}.
Nevertheless, the stochastic field dynamics -- defined by the Starobinsky Equation~(\ref{eq:starobinsky}) -- is an approximation to the underlying formulation in terms of Quantum Field Theory (QFT) that has remained somewhat obscure, despite previous works \cite{Tsamis:2005hd, Finelli:2008zg, Finelli:2010sh}.

Some questions that one might still raise in the stochastic formulation are:
\begin{itemize}
\item
The fact that the canonical commutator is much smaller than the anticommutator
is required for the self-consistency of the stochastic approach. This has been shown for the modes of linearized (non-interacting) fields \cite{Starobinsky:1986fx, Kiefer:1998qe} but can this be validated within a QFT calculation that systematically includes at least the leading non-linearities?

\item
How can the effective
stochastic approach be identified in terms of a truncation of the QFT calculation
at a certain order in an expansion parameter? Can this procedure be strictly justified?

\item
The separation into horizon-size and short-wavelength modes breaks de~Sitter invariance.
Can the stochastic results be confirmed in a framework that treats all field modes
on the same footing?
\end{itemize}

A powerful method to approach these questions is to set up the problem in
Euclidean de~Sitter space. For a massless scalar field with quartic interactions,
the leading infrared (IR) expansion for long-wavelength correlators corresponds to functionally
integrating over the constant mode ({\it i.e.} the zero mode) of the field. Since Euclidean
de~Sitter is compact, this reduces to a one-dimensional
integration~\cite{Rajaraman:2010xd}. It can also be observed that thus simplified functional
integrals coincide with the integrals over the probability distribution functions in
Stochastic Inflation, see Ref.~\cite{Beneke:2012kn}. Moreover, it has been demonstrated in \cite{Beneke:2012kn} that for Schwinger-Dyson equations derived from a two-particle-irreducible
effective action, the solutions for the two-point functions to leading IR order
take the form of free propagators with a dynamical mass. The resummation of an infinite
class of self-energy diagrams is then necessary in order to recover the results from
Stochastic Inflation or the functional integration of the constant mode.
Beyond the leading IR approximation, two-point functions for massive
scalar fields on Euclidean de~Sitter space to all orders in perturbation theory
have been investigated in Refs.~\cite{Hollands:2010pr,Marolf:2010nz}, where it is found that
these are well defined and that in particular, the field correlations exhibit
an exponentially decaying behaviour for large separations.
However, the decay at large Euclidean distances could not yet be proved for the massless case~\cite{Hollands:2011we}.
Therefore, calculations of leading IR effects in the massless
theory presently have to rely on the assumption that the correlation functions
computed without truncations are well-defined after all.

While the aforementioned investigations in Euclidean space
provide some substantial insights into interacting scalar field theory
on de~Sitter, calculations in a spacetime with Lorentzian signature,
as performed in the present work, are crucial to address the following
important points which are beyond the scope of Euclidean methods:
\begin{itemize}
\item
Depending on the initial conditions, the correlation functions
of light or massless fields exhibit a transient growth
or decay, {\it i.e.} the two point function evolves proportional to
the number of ${\rm e}$-folds, before reaching an ``equilibrium state''.
This can have important consequences on the evolution of light fields
during inflation and subsequently, in the Early Universe.
Clearly, this important feature requires calculations in the
Lorentzian spacetime.
\item
The analytic continuation between Euclidean and Lorentzian spacetimes requires
that the expansion of the latter is exactly of the de~Sitter form.
Cosmic inflation does however  break de~Sitter invariance, due to its definite
end along a spatial hypersurface, and potentially due to a beginning at a finite time.
Also while inflation takes place, there is a difference from de~Sitter expansion,
indicated by the observed deviation of the spectral index of the scalar power spectrum from
unity~\cite{Ade:2013zuv}.
\item
Even though there are theoretical arguments supporting the assertion that the late-time
limit of correlations in Lorentzian de~Sitter space can be obtained by analytic
continuation of the Euclidean results~\cite{Higuchi:2010xt}, it is desirable to demonstrate and to understand
this agreement
through a direct calculation of the Lorentzian observables within a QFT framework.
\end{itemize}

The appropriate tool for computing the time evolution of quantum correlation
functions is the Closed-Time-Path (CTP) formalism of QFT~\cite{Schwinger:1960qe, Keldysh:1964ud} (see {\it e.g.}~\cite{Jordan:1986ug} or~\cite{Kamenev:2009jj, Altland:2006si} for functional integral developments of the formalism and {\it e.g.}~\cite{Weinberg:2005vy, Prokopec:2002jn, Prokopec:2003tm, Seery:2007we, Seery:2007wf, Seery:2010kh} for cosmological applications), also known as the Schwinger-Keldysh or the ``in-in'' formalism. Since it involves two branches of dynamical evolution, forward and backward in time, we find it appropriate to also refer to it as Amphichronous QFT.

Perturbation theory on inflationary backgrounds with Amphichronous QFT can become complicated and, most importantly, breaks down at late times for (almost) massless fields due to the growth of individual terms in the series.
On the other hand, stochastic methods address both these problems because calculations in that framework are comparably simple, and they
offer a particularly convenient way of performing the necessary resummations~\cite{Tsamis:2005hd}. Thus, a refined picture of the stochastic dynamics and its relation to the underlying Lorentzian Amphichronous QFT is essential for understanding the behaviour of scalar fields during inflation.
In a number of computations reported in the literature,
various results for interacting Field Theory on de~Sitter
space in the presence of IR enhanced correlations are derived. Some of the different truncations
and also resummation strategies
that have been applied are {\it e.g.} usual
perturbation theory~\cite{Onemli:2002hr,Onemli:2004mb,Brunier:2004sb,Kahya:2006hc,Garbrecht:2011gu,Garbrecht:2013,Gautier:2013aoa,Tanaka:2013caa,Herranen:2013raa}, the Hartree approximation~\cite{Arai:2011dd,Arai:2012sh,Nacir:2013xca}, a $1/N$ expansion in
${\rm O}(N)$-symmetric theories~\cite{Riotto:2008mv,Serreau:2011fu,Prokopec:2011ms,Boyanovsky:2012nd,Serreau:2013psa,Gautier:2013aoa},
the Wigner-Weisskopf method~\cite{Boyanovsky:2012qs}, functional renormalization group
techniques~\cite{Burgess:2009bs} or other partial
resummation schemes~\cite{Youssef:2013by}. These approaches however do not appear to recover the
non-perturbative resummation that is readily performed in the stochastic approach.

Here, we therefore aim to elucidate the nature of the stochastic description by demonstrating its perturbative equivalence to the underlying QFT in the IR. For this
purpose, we choose the simple setting of a self interacting scalar field in de Sitter space, but our arguments should hold for more general inflationary spacetimes and Field Theories. Our results imply that indeed, stochastic inflation fully captures the leading IR behaviour of the underlying QFT and is safe for field theorists to use. From this we can
also conclude that the stochastic resummation
applies in the IR to QFT on de~Sitter space as well, thus proving a conjecture stated
{\it e.g.} in Refs.~\cite{Onemli:2002hr,Tsamis:2005hd}.

The outline of this paper is as follows: In Section~\ref{sec:stochastic_exp},
we review and present further details on the diagrammatic expansion of the correlation
functions from Stochastic Inflation that has been introduced in Refs.~\cite{Garbrecht:2013,Rigopoulos:2013exa}.
Turning to the Field Theory approach, the elementary propagators
as building blocks for Amphichronous Feynman diagrams in the Keldysh representation are presented in Section~\ref{sec:QFT-approach}.
The Feynman diagrams are then evaluated and brought to
a form that can be compared with their stochastic counterparts. The necessary approximations to leading IR order are carefully
justified, and this central part of the present work is presented in Section~\ref{sec:evaluationandcomparison}. Having demonstrated the perturbative equivalence between QFT and the stochastic approach in the IR, we then proceed in Section~\ref{sec:resum} to discuss the non-perturbative late-time limit of the stochastic dynamics where the field reaches an equilibrium state. We thus conclude that the non-perturbative equilibrium distribution that is attained by the field corresponds to the correct late-time resummation of the QFT series. This resummation also works for very light fields for which the QFT expansion fails completely at late times. We conclude in Section \ref{sec:discussion} by summarizing our results and touching upon the issue of de~Sitter invariance in the stochastic formalism. The notations and conventions used here are in line with Refs.~\cite{Garbrecht:2011gu,Garbrecht:2013}.

\section{Diagrammatic Expansion of Correlators in the \\
Stochastic Approach}\label{sec:stochastic_exp}

We consider a real scalar field in four spacetime
dimensions within the expanding
half of de~Sitter space, also known as the Poincar\'e patch, see
{\it e.g.} Ref.~\cite{Spradlin:2001pw} for a comprehensive discussion of the properties
and parameterizations of this spacetime.
The metric is therefore
fully parameterized through the value of the Hubble rate $H$.
Moreover, in comoving coordinates, it is manifest that the spatial sections are
flat such that we may seek spatially homogeneous solutions.
In terms of this coordinate choice, the stochastic
approach consists of separating long-wavelength
modes from short-wavelength ones.
Due to their IR enhancement,
the long modes can be treated as classical random variables
that can effectively be described by Langevin dynamics~\cite{Starobinsky:1986fx,Starobinsky:1994bd}, {\it i.e.} they obey the Starobinsky Equation

\beq
\dot{\phi} + \frac{\partial_\phi V}{3H} = \xi(t),
\label{eq:starobinsky}
\eeq

\noindent where $\xi(t)$ is a stochastic noise term that originates from
integrating out the
short-wavelength modes, and where
a dot denotes a derivative with respect to the comoving
time $t$.
The noise is Gau{\ss}ian and white, such that
it is fully determined by its two point correlation function, which reads
\beq
\la \xi(t)\xi(t^\prime) \ra = \frac{H^3}{4\pi^2}\delta(t-t^\prime)\,,
\label{eq:noise_correlation}
\eeq
where the expectation value $\la\dots\ra$ denotes the average over noise realizations. The field $\phi$ in Eq.~(\ref{eq:starobinsky}) should be understood as the average over a patch of physical size $\sim H^{-1}$ with the stochastic noise acting on each of these patches with practically zero correlation between different patches.

In order to cast the stochastic approach into a form that can be readily
connected with Field Theoretic elements we employ functional (path integral) techniques, making use of the formalism and the diagrammatic representation developed in Refs.~\cite{Garbrecht:2013,Rigopoulos:2013exa}. In the functional formulation of the Starobinsky Equation~(\ref{eq:starobinsky}), expectation values of any function $\mathcal{O}[\phi]$ w.r.t.~the different realizations of the stochastic force $\xi$ can be obtained from the following path integral

\beq\label{eq:stoch-path}
\langle \mathcal{O} [\phi] \rangle = 
\int D [\xi] {\rm e}^{- \frac{1}{2} \int dt \xi^2 \frac{4 \pi^2}{H^3}} \int D[\phi] \mathcal{O} [\phi] \delta (\dot{\phi} +\partial_{\phi} V / 3H - \xi)\mathcal{J}[\phi]\,,
\eeq

\noindent where $\mathcal{J}[\phi]$ is the Jacobian of the argument of the delta function with respect to the ``integration variable'' $\phi$: $\mathcal{J}[\phi]= \left|{\rm Det}\left[\frac{\delta}{\delta\phi}\left(\dot{\phi} +\partial_{\phi} V / 3H - \xi\right)\right]\right|$. The integration over the noise $\xi$ reflects the assumption that the latter is Gau{\ss}ian. To compute the determinant, we discretize the time interval in $N$ time steps of extent $\Delta t$ such that $\phi_i=\phi(t_i)$ and $\xi_i=\xi(t_i)$ with $i=0,\ldots , N$. Then the determinant can be written $\mathcal{J}=\left|{\rm Det}\mathcal{J}_{ij}\right|$, where

\beq
\mathcal{J}_{ij}=\frac{\partial}{\partial\phi_j}\left(\frac{\phi_i-\phi_{i-1}}{\Delta t}+\frac{\partial_\phi V(\phi_{i-1})}{3H}-\xi_{i-1}\right)\,.
\eeq

\noindent It is important to note that we have chosen a retarded regularization for the operator, {\it i.e.} all functions of $\phi$ and the stochastic force $\xi$ are calculated at the start of each timestep. With this choice, $\mathcal{J}_{ij}$ is $\frac{1}{\Delta t}$ on the diagonal and the only other non-zero elements are the $\mathcal{J}_{i,i-1}$ entries. Absorbing the $\frac{1}{\Delta t}$ factors in the measure, we see that $\mathcal{J}[\phi]=1$. Note that, with the appropriate normalization of the measure $D[\xi]$, the integral is a realization of the identity: $ \la \mathbb{I} \ra = 1$. Nevertheless, it is convenient to make use of the partition function
\beq
\mathcal{Z} \equiv \int D [\xi] {\rm e}^{- \frac{1}{2} \int dt \xi^2 \frac{4 \pi^2}{H^3}} \int D[\phi] \delta (\dot{\phi} +\partial_{\phi} V / 3H - \xi) = 1\,,
\eeq
then to introduce an auxiliary field $\psi$ in order to replace the $\delta$ function
with a functional Fourier integral and eventually to integrate out the noise $\xi$. The partition function then reads\footnote{Here, the auxiliary field
$\psi$ has a different mass dimension than in previous work~\cite{Garbrecht:2013,Rigopoulos:2013exa}, such that the propagators $G^{R,A}$ and $F$ all have mass dimension {\it two}.}

\beq\label{eq:partition 1}
 \mathcal{Z} = \int D [\phi] D [\psi] {\rm e}^{- \int dt \left\{\frac{{\rm i}}{H^2} \psi \left(\dot{\phi} + \frac{\partial_{\phi} V}{3H} \right) + \frac{1}{8 \pi^2 H} \psi^2\right\}}\,.
\eeq

\noindent Adding couplings $ -{\rm i}\int dt J_\psi\psi$ and ${-\rm i}\int dt J_\phi\phi$ with currents in the exponent defines $\mathcal{Z}[J_\psi,J_\phi]$ from which $n$-point functions can be computed in the usual way by taking derivatives with respect to $-{\rm i}J_\psi$ and $-{\rm i}J_\phi$. Note that, unlike the standard QFT partition function, $\mathcal{Z}[0,0]=1$, and expectation values can be obtained directly from $\mathcal{Z}$ without receiving
multiplicative contributions from vacuum bubbles.

In order to prepare for a perturbative expansion, we replace
$\dot\phi\psi \to \frac12( \dot{\phi}\psi - \phi\dot{\psi})$ under the
integral, and we
decompose the potential as $V = \frac{1}{2}m^2\phi^2 + V_{\mathrm{int}}(\phi)$. This
yields

\bea
\label{def:go1}
 \mathcal{Z} &=& \int D [\phi] D [\psi] {\rm e}^{- \mathrm{i}\int dt \left\{ \frac{1}{2} \bmatr \phi, & \psi \ematr \bmatr 0 & \frac{1}{H^2}(-\partial_t + \frac{m^2}{3H}) \\  \frac{1}{H^2}(\partial_t + \frac{m^2}{3H}) & -\frac{i}{4 \pi^2 H} \ematr \bmatr \phi \\ \psi \ematr + \frac{\partial_\phi V_{\mathrm{int}}}{3H^3}\psi\right\}}
 \\
 \notag
 &\equiv& \int D [\phi] D [\psi] {\rm e}^{- \mathrm{i}\int dt \left\{ \frac{1}{2} \bmatr \phi, & \psi \ematr \mathbb{G}_0^{-1} \bmatr \phi \\ \psi \ematr + \frac{\partial_\phi V_{\mathrm{int}}}{3H^3}\psi\right\}},
\eea

\noindent where the last equality defines $\mathbb{G}_0^{-1}$. The latter is the functional and matrix inverse of the free propagator $\mathbb{G}_0$, that is given by

\beq
\mathbb{G}_0(t,t^\prime) = \bmatr \la\phi(t)\phi(t^\prime) \ra & \la \phi(t)\psi(t^\prime) \ra \\ \la\psi(t)\phi(t^\prime) \ra & \la \psi(t)\psi(t^\prime) \ra  \ematr \equiv \bmatr F(t,t^\prime) & -\mathrm{i}G^R(t,t^\prime) \\ -\mathrm{i}G^A(t,t^\prime) & 0  \ematr\,,
\label{def:go}
\eeq

\noindent where the above equality defines the free propagators $G^{R,A}$ and $F$. It should be emphasized that the null entry in $\mathbb{G}_0$ is a direct consequence of the definition of $\mathbb{G}_0^{-1}$.
It occurs due to the fact that $\psi$ is an auxiliary field and therefore not dynamical. Using Eqs.~(\ref{def:go1},\ref{def:go}) and the relation $\mathbb{G}_0 \star \mathbb{G}_0^{-1} (t,t^\prime) = \mathbb{I}_{2\mathrm{x}2}\delta(t,t^\prime)$,
we observe that $G^{R,A}$ are the retarded and advanced propagators for the operator $\frac{1}{H^2}(\partial_t + \frac{m^2}{3H})$, while the statistical correlator is the two-point function of the original field  $F(t,t^\prime) = \la\phi(t)\phi(t^\prime) \ra$.
These Green functions can be easily found, and they read
\begin{subequations}
\label{free:propagators}
\bea
G^R(t,t^\prime) &=& G^A(t^\prime,t) = H^2 {\rm e}^{-\frac{m^2}{3H}(t-t^\prime)}\Theta(t-t^\prime), \label{def:gret_stoc} \\
F(t,t^\prime) &=& \frac{3H^4}{8\pi m^2}\left({\rm e}^{-\frac{m^2}{3H}|t-t^\prime|} - {\rm e}^{-\frac{m^2}{3H}(t+t^\prime)} \label{def:F_stoc} \right)\,,
\eea
\end{subequations}

\noindent
where we have imposed the initial condition $F(0,0)=0$.
Before going on, let us remark that $F(t,t) \approx \frac{H^3}{4\pi}t$ for $m \to 0$. This famous secular behaviour~\cite{Vilenkin:1982wt, Linde:1982uu, Vilenkin:1983xp, Tsamis:1993ub} signals the breakdown of the perturbative expansion for large enough times and masses that are small ({\it e.g.} $m^2\ll\sqrt\lambda H^2$ for the quartic interaction assumed below)
or vanishing. In order to maintain $F(t,t)$ regular also for infinitely late times,
we first restrict
to the case where the squared mass is strictly positive. For ${\rm min}(Ht,Ht')\gg (3H^2)/2m^2$, {\it i.e.} assuming that the stochastic process (equivalently inflation) has started early enough,
the growth of the equal-time correlations
saturates, such that $F(t,t^\prime)$ only depends on the time separation $|t-t^\prime|$:
\beq
F(t,t^\prime) = \frac{3H^4}{8\pi^2 m^2}{\rm e}^{-\frac{m^2}{3H}|t-t^\prime|}.
\label{eq:F_late}
\eeq
\noindent
In Section~\ref{sec:resum},
we comment on the important case $m^2=0$.

\begin{figure}
\begin{center}
\parbox{3.1cm}
{
\center
\vskip1.25cm
\epsfig{file=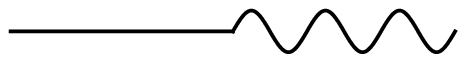,scale=0.6}

$-{\rm i}G^R(t,t')$
}
\parbox{3.1cm}
{
\center
\vskip1.25cm
\epsfig{file=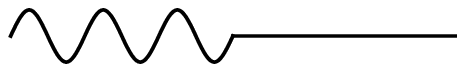,scale=0.6}

$-{\rm i}G^A(t,t')$
}
\parbox{3.1cm}
{
\center
\vskip1.35cm
\epsfig{file=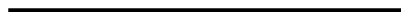,scale=0.6}

\vskip0.1cm
$F(t,t')$
}
\parbox{3.1cm}
{
\center
\vskip1.25cm
\epsfig{file=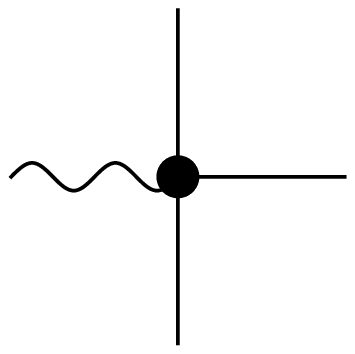,scale=0.6}

$-{\rm i}\frac{\lambda}{3H^3}\int d\tau$
}
\end{center}
\caption{The elements out of which stochastic diagrams are constructed. The choice of vertex factor implies that the assembled diagrams should be divided by their symmetry factor.}
\label{fig:stochastic-propagators}
\end{figure}

The partition function in its form~(\ref{def:go1}) readily leads to a diagrammatic expansion for the
correlation functions, using the free propagators~(\ref{free:propagators}) as internal lines. The vertices
derive from the interaction potential $V_{\mathrm{int}}$, for which we take
a quartic coupling
$V_{\mathrm{int}} = \frac{\lambda}{4!}\phi^4$, such that
the mixing term in Eq.~(\ref{def:go1}) becomes

\beq
 \frac{\partial_\phi V_{\mathrm{int}}}{3H^3}\psi = \frac{\lambda}{3!}\frac{\psi\phi^3}{3H^3}.
\label{def:vertex}
\eeq
\noindent
Putting these elements together, we can derive a set of Feynman rules that is presented
in Figure~\ref{fig:stochastic-propagators} ({\it cf.} Ref.~\cite{Garbrecht:2013}).

Note that the ``symmetric'' interaction term $\psi^3\phi$ is absent in the stochastic description. This is in contrast to the corresponding QFT Feynman rules in
the Keldysh representation which contain both $\psi^3\phi$ and $\psi\phi^3$ vertices - {\it cf.} Eq.~(\ref{action:Keldysh}) below.
However, the $\psi^3\phi$ vertex is irrelevant for long wavelengths due to the IR enhancement of the field correlations: While the statistical propagator~(\ref{def:F_stoc}) is enhanced in the IR, the retarded and advanced propagators~(\ref{def:gret_stoc}) remain regular in the limit $m^2/H^2\to0$. QFT diagrams with vertices connecting
with three retarded or advanced propagators are therefore suppressed in powers of $m^2/H^2$ compared to diagrams where only one of these propagators connects to each vertex. The irrelevance of the $\psi^3\phi$ vertex can be understood as a consequence of the fact that the long modes behave classically. In the stochastic approach this vertex is therefore absent by construction.

\section{Field Theoretical Approach} \label{sec:QFT-approach}

The agreement of the Field Theoretical and stochastic calculations for the 2-point function was previously demonstrated in Ref.~\cite{Garbrecht:2013}
up to order $\lambda^2$, a result relevant in the regime where $H^2\gg m^2\gg\sqrt\lambda H^2$, such that a perturbative truncation at some order in $\lambda$ is meaningful.
In that work, the QFT calculation
has been performed in the basis defined by the $\pm$ branches of the CTP. We refer to this basis as the $\pm$ basis or the Wightman basis because the Wightman propagators appear there
explicitly as the Green functions connecting the $+$ with the $-$ branch of the Amphichronous time evolution or vice versa.
For the present purposes, it is however advantageous to work in the Keldysh basis~\cite{Keldysh:1964ud} instead,
where the retarded, advanced and statistical Green functions appear as the elementary propagators. The reason is twofold: a) It can readily
be noted from the calculation in Ref.~\cite{Garbrecht:2013} that the expressions for the loop integrals are best arranged in terms of the Keldysh propagators. b) The structure of the free stochastic Green function~(\ref{def:go}) from Section~\ref{sec:stochastic_exp} emerges naturally within the Keldysh basis, making it a preferred choice when comparing between the stochastic and Field Theoretic formalisms. In the following, to facilitate a comparison with Ref.~\cite{Garbrecht:2013}, we review how to transform between the Wightman and the Keldysh bases.

In the Wightman basis, the field is divided into $\phi_+$ and $\phi_-$, corresponding to the forward and backward
branches of the Amphichronous time-evolution. The free propagator ${\rm i} \Delta$ satisfies the Klein-Gordon equation
\begin{align}
\label{KG:free}
a^4\left(\nabla_x^2-m^2_{}\right){\rm i}\Delta^{(0)fg}(x;x^\prime)
=fg \delta^{fg}\,{\rm i}\delta^4(x-x^\prime)
\,,
\end{align}
where $f,g=\pm$ are CTP indices and $(\nabla_x)_\mu$ is the covariant derivative with respect to $x$. The
transformation from the Wightman to the Keldysh basis
is performed with the matrix
\begin{align}
U = \frac{1}{\sqrt{2}} \bmatr 1 & 1\\ 1 & -1\ematr\,,
\end{align}
that acts on the field components as follows:

\beq
\bmatr \phi \\ \psi \ematr = U \cdot \bmatr \phi_+ \\ \phi_- \ematr =  \frac{1}{\sqrt{2}}  \bmatr  \phi_+ + \phi_-\\ \phi_+ - \phi_- \ematr\,.
\eeq

\noindent For the free propagators, this implies the relation

\begin{align}
\label{eq:transformation}
U.\bmatr{\rm i} \Delta^{T} \;&\; {\rm i} \Delta^{<} \\
   {\rm i} \Delta^{>}\; & \;{\rm i} \Delta^{\bar{T}} \ematr .U^\dagger =& \bmatr {\rm i} \Delta^{<}+{\rm i} \Delta^{>} \;&\; {\rm i}\Delta^{T} - {\rm i} \Delta^{<} \\
											      {\rm i} \Delta^{T} - {\rm i} \Delta^{>} \;& \;0 \ematr
\\ 
\equiv&
											      \bmatr F (x, x')\; &\; -i G^{R} (x,x') \\
																		-i G^{A} (x,x') & 0 \ematr,\label{eq:transformation2}
\end{align}
where the last equality defines
the statistical propagator $F(x,x^\prime)$ as well as the causal ones $G^{R,A}(x,x^\prime)$. For the first equality, we have used that
\begin{align}
 {\rm i} \Delta^{T}+{\rm i} \Delta^{\bar{T}}
 = {\rm i} \Delta^{<}+ {\rm i} \Delta^{>}
 \,.
\end{align}
We employ here the same symbol for the QFT propagators on de~Sitter space as for their counterparts in the stochastic formalism,
defined in Section~\ref{sec:stochastic_exp}. Eq.~(\ref{eq:transformation2}) should then be compared with Eq.~(\ref{def:go}),
which both share a similar structure. It should be clear however
that 
these quantities are intrinsically different.
It is particularly important to note that
\begin{itemize}
\item
in the stochastic approach, $\phi$ is a classically stochastic random
variable while in the Field Theoretical approach, it refers to a field
operator,
\item
the field $\psi$ is here fully dynamical, in contrast to its stochastic counterpart,
which is an auxiliary field.
\end{itemize}
However, as we will see eventually, the stochastic and QFT fields can be identified in the IR regime.


The solutions to Eq.~(\ref{KG:free}) are well known and can be expressed exactly
in terms of hypergeometric functions.
In Ref.~\cite{Garbrecht:2013}, these are then
expanded to leading order in IR-enhancement $H^2/m^2$ and in large
separations $|y|$,
where the distance function is given by~\cite{Onemli:2002hr,Tsamis:1992xa}
\begin{align}
\label{invariant:length:y}
y(x;x^\prime)=\frac{(\eta-\eta^\prime)^2-(\mathbf x-\mathbf x^\prime)^2}{\eta\eta^\prime}
\,,
\end{align}
which we express in terms of the conformal time $\eta$ that is related to comoving time as
$\eta = - \frac{1}{a_0 H} {\rm e}^{-H t}$, where $a_0$ is a constant.
When transformed
into the Keldysh basis, these results take the form
\begin{subequations}
\label{KeldyshGF}
\begin{align} \label{KeldyshRA}
-{\rm i} G^{R, A} (x,x') = {\rm i} \Delta^{(0)R, A} (x,x') = \frac{H^2}{4 \pi^2} \left( - \frac{\rm i}{2} \right) \arg y^{R, A} |y|^{-\frac{m^2}{3 H^2}},
\\
\label{KeldyshF}
F (x, x') = {\rm i} \Delta^{(0)>} (x, x') + \Delta^{(0)<} (x, x') = \frac{3H^4}{4 \pi^2 m^2} |y|^{-\frac{m^2}{3 H^2}},
\end{align}
\end{subequations}
where the argument of $y$ can be expressed as
\beq
\arg y^{R} (x, x') = \arg y^{A} (x', x) = 2 \pi \vartheta (\eta - \eta ') \vartheta ((\eta - \eta ')^2 - ({\mathbf x} - {\mathbf x}')^2)\,,
\eeq
which follows from the appropriate boundary conditions for the Green functions. The
divergence in the statistical propagator $F(x,x^\prime)$ for $m\to 0$ is due to the fact
that for a minimally coupled and massless free scalar field, there is no de~Sitter invariant
quantum state~\cite{Allen:1985ux,Allen:1987tz}.

This form for the propagators is valid only for a light scalar field ($m^2 \ll H^2$).
According to Eqs.~(\ref{KeldyshRA}) and~(\ref{KeldyshF}), the propagators decay for large separations\footnote{This can be interpreted as the same physical effect that leads to a
blue-tilted power-spectrum of inflationary perturbations from a massive scalar field due to the $\eta$ term.}, {\it i.e.} for large values of $|y|$. However, as the mass $m \rightarrow 0$, {\it i.e.} for  $m^2\ll\sqrt\lambda H^2$, the decay of the IR fluctuations can become slow enough for the perturbation expansion to break down.
Note also that the statistical propagator has a factor $H^2/m^2$ of IR enhancement. In contrast, the retarded and advanced propagators that encompass the spectral information for the field excitations remain regular for $m^2/H^2\to 0$. This is in accordance with the corresponding stochastic two-point functions given in Eq.~(\ref{free:propagators}).

For the diagrammatic expansion we need the Feynman rules for the vertices that are
connected by the propagators discussed above. To obtain them, we start from the Amphichronous effective action
\beq
S[\phi_+, \phi_-] = \int d^4 x [\mathcal{L} (\phi_+) - \mathcal{L} (\phi_-)]\,,
\eeq
which reads in the Keldysh basis
\beq
\label{action:Keldysh}
S[\phi_+, \phi_-] = \int d^4 x \,\, \sqrt{-g} [\frac{1}{2} (\psi \hat{\mathcal{O}} \phi + \phi \hat{\mathcal{O}} \psi) - \frac{1}{2} \frac{\lambda}{3 !} (\phi^3 \psi + \phi \psi^3)]\,,
\eeq
with the kinetic operator
\beq
\hat{\mathcal{O}} = \nabla^2 + m^2\,.
\eeq

Notice that there are two types of vertices: $\phi \psi^3$ connects with at least three retarded or advanced propagators, and $\phi^3 \psi$ connects with at least one retarded or advanced propagator. It can therefore readily be seen that, due to the smaller degree of IR enhancement, diagrams containing a $\phi \psi^3$
vertex can be neglected at leading order in the IR enhancement.
This is clearly reflected in the stochastic diagrammatic expansion, where the $\phi \psi^3$ is absent in first place, {\it cf.} the discussion
at the end of Section~\ref{sec:stochastic_exp} and also in Ref.~\cite{Rigopoulos:2013exa}.

\begin{figure}
\begin{center}
\parbox{3.0cm}
{
\center
\vskip1.25cm
\epsfig{file=Ret.eps,scale=0.6}

$-{\rm i}G^R(x,x')$
}
\parbox{3.0cm}
{
\center
\vskip1.25cm
\epsfig{file=Adv.eps,scale=0.6}

$-{\rm i}G^A(x,x')$
}
\parbox{3.0cm}
{
\center
\vskip1.35cm
\epsfig{file=F.eps,scale=0.6}

\vskip0.1cm
$F(x,x')$
}
\parbox{3.0cm}
{
\center
\vskip1.25cm
\epsfig{file=SVertex.eps,scale=0.6}

$-{\rm i}\frac{\lambda}{2}\int d^4 x \,\,a^4(x)$
}
\parbox{3.0cm}
{
\center
\vskip1.25cm
\epsfig{file=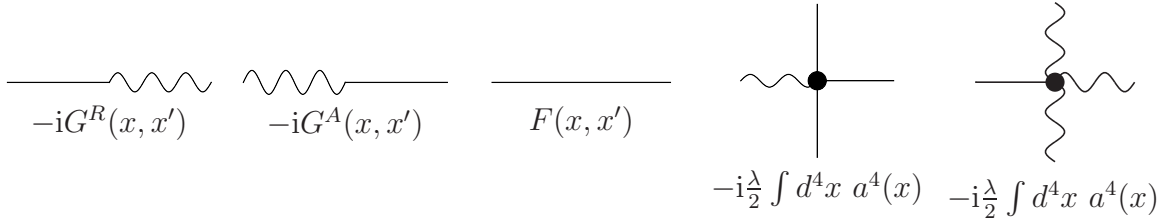,scale=0.6}

$-{\rm i}\frac{\lambda}{2}\int d^4 x \,\,a^4(x)$
}
\end{center}
\caption{The elements out of which the QFT Feynman diagrams are constructed. The choice of vertex factor implies that the assembled diagrams should be divided by their symmetry factor. Note the similarity with the stochastic diagrams of Section \ref{sec:stochastic_exp} as well as the obvious differences: Integrations extend over all of spacetime and the extra $\psi^3\phi$ vertex appears.
}
 \label{fig:QFT-propagators}

\end{figure}

We show the Feynman rules for QFT in the Keldysh basis in Figure~\ref{fig:QFT-propagators}. The graphical similarities with the diagrammatic expansion of the stochastic theory are evident, as are the obvious differences: There is an extra vertex and integrations extend over the whole spacetime instead of temporal integrations only. With these rules one can formally express correlation functions of $\phi$ to all orders in the loop expansion, but of course, an exact evaluation of all convolution integrals appears not to be practicable. In the next Section, we show however that to leading IR approximation, the spatial integrals can be performed, establishing an agreement with the stochastic diagrams of Section~\ref{sec:stochastic_exp}.

\section{Equivalence between Field Theoretical and Stochastic Diagrams at
Leading IR Order}
\label{sec:evaluationandcomparison}

In this Section, we explicitly demonstrate that the QFT diagrams calculated
from the Feynman rules of Figure~\ref{fig:QFT-propagators}
evaluate to the same results at leading IR order as the corresponding stochastic diagrams constructed from Figure~\ref{fig:stochastic-propagators}. Since the extra $\psi^3\phi$ vertex is irrelevant in the IR, the main difference is that the stochastic propagators~(\ref{free:propagators}) are purely time-dependent, which is a consequence of the separation into short and long modes, that abandons manifest de~Sitter invariance. In contrast, the QFT propagators depend also on the spatial coordinates, and they are de~Sitter invariant, which also holds true for their approximate forms~(\ref{KeldyshRA}) and (\ref{KeldyshF}).

In the following, we therefore establish the agreement between the diagrams by
performing the spatial integrals to leading IR approximation, thus abandoning manifest de~Sitter invariance as well. For this purpose, we first
show how the agreement is achieved when treating the factors of the
propagators that depend on the spatial separations as approximately constant
to leading IR order.
Then, we justify that the contributions from the integration regions where the
latter approximation is not valid are exponentially small (in the parameter $H^2/m^2$)
and therefore negligible.

\paragraph{Reduction of QFT Diagrams to the Stochastic Form. --}

Since the stochastic propagators are purely time-dependent, we first separate the space dependence from the
QFT propagators within an explicit factor. The causal and the statistical propagators~(\ref{KeldyshRA}) and~(\ref{KeldyshF}) share the same non-trivial dependence on the de~Sitter invariant length $y(x,x^\prime)$, which is raised to
a power that determines the leading IR behavior at large space- or time-like separations.
We rewrite these terms as
\beq \label{power_term}
|y|^{-\frac{m^2}{3 H^2}} = \left| \frac{\eta \eta'}{(\eta - \eta')^2 - (\mathbf{x} - \mathbf{x}')^2} \right|^{\frac{m^2}{3 H^2}} = \left( \frac{\eta \eta'}{(\eta - \eta')^2} \right)^{\frac{m^2}{3 H^2}} \left| \frac{1}{1 - \delta^2} \right|^{\frac{m^2}{3 H^2}},
\eeq
where $\delta^2 = \frac{(\mathbf{x} - \mathbf{x}')^2}{(\eta - \eta')^2}$. As it
is shown in Ref.~\cite{Garbrecht:2013}, the integration over regions with
large time separations ({\it i.e.} $|t-t'| \gg 1/H$) accumulates IR-enhancement factors
$\sim\frac{H^2}{m^2}$ in addition to the explicit factors present in the statistical
propagators. In these integration regions, based on the approximation
\begin{align}
\frac{\eta \eta'}{(\eta - \eta')^2}
=\frac{{\rm e}^{-H|t-t^\prime|}}{1+{\rm e}^{-2H|t-t^\prime|}-2{\rm e}^{-H(t+t^\prime)}}
\approx {\rm e}^{-H|t-t^\prime|}
\,,
\end{align}
we can replace the time-dependent factor
in Eq.~(\ref{power_term}) with
${\rm e}^{- \frac{m^2}{3 H} |t - t'|}$. Moreover, since $m^2 \ll H^2$, we can expand
\begin{align}
\label{IR:approximation}
\left| \frac{1}{1 - \delta^2} \right|^{\frac{m^2}{3 H^2}} = 1 +{\cal O}\left(\frac{m^2}{H^2}\right)\,,
\end{align}
for separations satisfying
\beq \label{approx_valid}
1-\delta^2 \gg  \exp \left(- \frac{3 H^2}{m^2}\right) \quad {\textnormal{and}} \quad  \exp \left(- \frac{3 H^2}{m^2}\right) \ll \delta^2-1 \ll\exp \left( \frac{3 H^2}{m^2}\right).
\eeq

Since $\delta^2 \in [0,+\infty]$ for any physical separation, one observes that the larger the IR enhancement, {\it i.e} the closer $m$ is to zero,
the wider is the window of validity~(\ref{approx_valid}) for the approximation~(\ref{IR:approximation}). Then, for separations satisfying the above conditions, the power term (\ref{power_term}) is approximated by
\beq \label{approximation}
|y|^{-\frac{m^2}{3 H^2}} = {\rm e}^{- \frac{m^2}{3 H} |t - t'|}
\times\left(1+{\cal O}\left(\frac{m^2}{H^2}\right)\right)\,,
\eeq
such that we can reexpress the Field Theoretical propagators as
\beq \label{approxmated F}
F(x, x') = \frac{3H^4}{4 \pi^2 m^2} {\rm e}^{-\frac{m^2}{3 H} |t - t'|},
\eeq
and
\beq
G^{R} (x, x') = G^A (x', x) = \frac{H^2}{4 \pi} \theta(t-t') \theta((\eta - \eta ')^2 - ({\mathbf x} - {\mathbf x'})^2) {\rm e}^{-\frac{m^2}{3 H} (t - t')}.
\eeq
We note that up to a factor of 2, Eq.~(\ref{approxmated F}) is of the same form as the statistical propagator $F(t, t')$ in the stochastic approach~(\ref{eq:F_late}); in particular it is space independent.
Moreover, in diagrams contributing to leading IR order,
each of the vertex integrals involves only one retarded or advanced propagator. These contribute
a factor $\frac{1}{4 \pi} \theta((\eta - \eta')^2 - ({\mathbf x} - {\mathbf x'})^2)$ that we absorb into the vertex integral,
\beq
\label{approximated_lambda}
- {\rm i} \frac{\lambda}{2} \int d^4 x' a^4 (x')  \frac{1}{4 \pi} \theta((\eta - \eta')^2 - ({\mathbf x} - {\mathbf x'})^2) = - {\rm i} \frac{\lambda}{6} \int d \eta ' \frac{(\eta - \eta')^3}{H^4 \eta '^4} \approx - {\rm i} \frac{\lambda}{6 H^3} \int d t'
\,.
\eeq
The accuracy of the latter approximation is to be understood in the sense that when integrating over a function $\sim\vartheta(t-t^\prime)\exp(-\alpha \frac{m^2}{3H}|t-t^\prime|)$, there is a correction ${\cal O}(\alpha \frac{m^2}{3H})$, which is
negligible to leading IR order.

The above integration can be done for each individual vertex in a Feynman diagram contributing to leading IR order,
{\it i.e.} a diagram where each vertex connects to precisely one causal propagator.
Hence, we can effectively replace the retarded and advanced propagators with
\beq
 G^{R} (x, x')\,, G^A (x', x) \to \theta(t-t') H^2 {\rm e}^{-\frac{m^2}{3 H} (t - t')}\,,
 \label{approximated_gr}
\eeq

\noindent such that the effective vertex contribution is now given by  $- {\rm i} \frac{\lambda}{6 H^3} \int d t'$.
We stress that both the effective vertex and the effective causal propagators are only valid when evaluated
under the convolution integrals that arise from Feynman diagrams at leading IR order.
Their form is almost identical to the corresponding quantities in the
 stochastic formulation, {\it cf.} Eqs.~(\ref{def:gret_stoc}) and~(\ref{def:vertex}).
 The only remaining differences compared with the stochastic calculation are a factor of $1/2$ in the vertex coefficients as well as a factor 2 in the statistical correlations~(\ref{approxmated F}) and (\ref{eq:F_late}).
 We see that these discrepancies compensate when rescaling the QFT fields in the Keldysh basis $\phi \rightarrow \phi/\sqrt{2}$ and $\psi \rightarrow \sqrt{2}\psi$. Then, ignoring the $\psi^3\phi$ vertex, $\phi$ becomes equivalent to the average field between the forward and backward time contours $\phi\rightarrow \left(\phi_++\phi_-\right)/2$ while $\psi$ now corresponds to the auxiliary field of the stochastic formalism. $G^R$ and $G^A$ remain unaffected while $F$ and the vertex $\psi\phi^3$ now coincide with the expressions from the stochastic approach.


\paragraph{Contributions Close to the Light Cone and at Large Spatial Separations. --}

We have demonstrated above that
the truncation of the series~(\ref{IR:approximation}) at leading order readily leads to QFT Feynman diagrams
that are identical to the stochastic ones. Next, we estimate the contributions from those regions of integration where
this expansion breaks down and show that these are negligible to leading IR order. This is necessary because
in any generic Feynman diagram the convolution integral runs over the entire Poincar\'e patch and hence also receives contributions from separations
where the approximation~(\ref{IR:approximation}) is not valid.
From the criteria~(\ref{approx_valid}), we can categorize these separations into
a) regions in the vicinity of the light cone:
\beq \label{vici_ligcon}
1 - \alpha \exp \left(- \frac{3 H^2}{m^2}\right) < \delta^2 < 1 + \alpha \exp \left(- \frac{3 H^2}{m^2}\right)
\eeq
and b) large space-like separations, {\it i.e.}, large $\delta^2$:
\beq \label{large_delta}
\delta^2 > 1 + \frac{1}{\alpha} \exp \left(\frac{3 H^2}{m^2}\right)\,,
\eeq
where $1 \ll \alpha \ll \exp (\frac{3 H^2}{m^2})$ is a constant.

We aim to show that the regions specified above are exponentially small and therefore only
lead to contributions to the Feynman diagrams that are negligible at leading IR order.
While it is somewhat obvious that region a) corresponds to an exponentially small restriction of the
integration volume in the directions perpendicular to the light cone,
we nonetheless work this out explicitly in the following. The identification of region b)
turns out to be slightly less straightforward because we need to make use of the causal structure
of the vertex integrals in a Feynman diagram.

In a Feynman diagram contributing at leading IR order
({\it i.e.} all vertices are of the $\phi^3 \psi$ type), the vertex integrations are of the general form
\bea \label{vertex_int}
I(\{x_i\}) &=& -{\rm i}\frac{\lambda}{2}\int d^4 x^\prime a^4(x^\prime)\frac{H^2}{4\pi} \vartheta(\eta^\prime-\eta_1)
\vartheta((\eta^\prime-\eta_1)^2-(\mathbf{x}^\prime-\mathbf{x}_1)^2) \nonumber \\
&  \times & |y(x^\prime,x_1)|^{-\frac{m^2}{3H^2}}
|y(x^\prime,x_2)|^{-\frac{m^2}{3H^2}}
|y(x^\prime,x_3)|^{-\frac{m^2}{3H^2}}
|y(x^\prime,x_4)|^{-\frac{m^2}{3H^2}}\,.
\eea
Here the vertex at $x'$ connects with another vertex or an external point $x_1$ through a causal propagator and with three additional points through statistical propagators.
While it does not appear possible to exactly evaluate this integral
analytically, it is straightforward to estimate ({\it cf.} Ref.~\cite{Garbrecht:2013})
\begin{align}
\label{I:estimate}
I(\{x_i\})\sim \frac{\lambda}{m^2}\prod\limits_{i=2,3,4}\left(\frac{\eta_i}{\eta_1}\right)^{-\epsilon}\,,
\end{align}
where $\epsilon = \frac{m^2}{3H^2}$ and the IR enhancement is reflected by the divergence for $m\to 0$. For large ratios
$\eta_i/\eta_1$ there is an extra suppression because of the decay of correlations with
very early times.

The power terms can be rewritten in the form of Eq.~(\ref{power_term}) by introducing $\delta_i^2 = \frac{(\mathbf{x}_i - \mathbf{x}')^2}{(\eta_i - \eta')^2}$ where $i = 1,2,3,4$. The spatial integration in Eq.~(\ref{vertex_int}) then takes the form
\beq \label{spatial_int}
\int d^3 \mathbf{x}' \vartheta((\eta^\prime-\eta_1)^2-(\mathbf{x}^\prime-\mathbf{x}_1)^2) \prod_{i = 1}^4 \left| \frac{1}{\tilde{\delta}_i^2} \right|^\epsilon = 4 \pi (\eta_1 - \eta')^3 \int_0^1 d |\tilde{\delta}_1| \delta_1^2 \prod_{i = 1}^4 \left| \frac{1}{\tilde{\delta}_i^2} \right|^\epsilon,
\eeq
where $\tilde{\delta}_i^2 = 1 - \delta_i^2$. 
When all separations $x^\prime-x_i$ comply with the condition~(\ref{approx_valid}), we approximately take $| {1}/{\tilde{\delta}_i^2}|^\epsilon \approx 1$, and the term $\frac{4 \pi}{3} (\eta_1 - \eta')^3$ in Eq.~(\ref{spatial_int}) can be passed to the temporal integration according to Eq.~(\ref{approximated_lambda}) -- see the discussion above.

Now consider an integration region where $\delta^2_j$ lies in the vicinity of the light cone, {\it i.e.} within the region~(\ref{vici_ligcon}), while the remaining $\delta^2_i$ satisfy the relation~(\ref{approx_valid}) and can therefore
be approximated by $\tilde\delta_i^2\approx 1$ for $i\not=j$.
Then,
using $|\tilde{\delta}_j|$ as a variable of integration, we obtain from the region defined by relation~(\ref{vici_ligcon})
the contribution
\beq
\label{I:spatial}
2\int_0^{\frac{1}{2} \alpha {\rm e}^{- 1/ \epsilon}} d |\tilde{\delta}_j| \delta_j^2 \left| \frac{1}{\tilde{\delta}_j} \right|^{2  \epsilon} = \left. \frac{2 \tilde{\delta}_j^{1 - 2 \epsilon}}{1 - 2 \epsilon} \right|_0^{\frac{1}{2} \alpha {\rm e}^{- 1/ \epsilon}}\,,
\eeq
where we have used that $\delta_j\approx 1$ close to the light cone.
Therefore, compared to the contribution to the integral
from the regime~(\ref{approx_valid}) [or, more directly, the corresponding
factor in Eq.~(\ref{I:estimate}), which is of order one], the piece from the vicinity of the light cone is exponentially suppressed by a factor $\alpha \exp({- \frac{3 H^2}{m^2}})$.
It should be clear that similar arguments apply when more than one of the separations in the integral~(\ref{vertex_int})
are close to the light cone simultaneously because the regions where this
may occur are only exponentially small and only contain integrable singularities for $\epsilon\ll 1$.
We should also recall that while the region~(\ref{vici_ligcon}) is exponentially
narrow, the leading inaccuracy of our approximations is of order $m^2/H^2$ as stated through Eq.~(\ref{IR:approximation}).

\begin{figure}
  \centering
  \includegraphics[width=8cm]{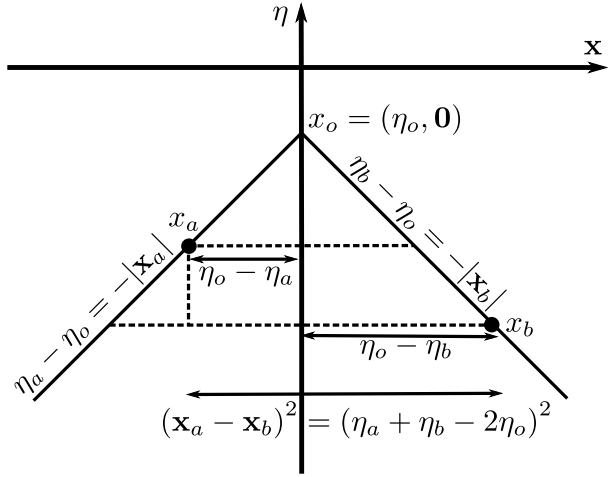}\\
  \caption{For a correlation function $\langle \phi(x_o)\phi(x_o) \rangle$, the support of the spacetime integration is
  given by the past light cone of $x_o$ because of causality. The diagram illustrates
  that this implies a maximal spatial distance for any two points with given $\eta_a$ and $\eta_b$. }\label{fig:lightcone}
\end{figure}

Now we turn to large space-like separations defined by relation~(\ref{large_delta}).
Since the long-wavelength fluctuations vary by definition only very slowly,
these observables are obtained by evaluating the coincident correlation function
$\langle\phi(x_o) \phi(x_o)\rangle$ and subtracting the short-wavelength contributions
that are independent from the background expansion. We choose for simplicity
$x_o=(\eta_o,\mathbf{0})$. Now, due to causality, all points that contribute
to the Feynman diagrams must lie within the past light cone of $x_o$, a fact we
illustrate in Figure~\ref{fig:lightcone}. This implies that
all two-point separations that occur in the generic integral~(\ref{vertex_int})
satisfy the bound (we identify $x^\prime=x_a$ and $x_i=x_b$ in Figure~\ref{fig:lightcone})
\begin{align}
\delta^2 (x^\prime, x_i) = \frac{(\mathbf{x}^\prime - \mathbf{x}_i)^2}{(\eta^\prime - \eta_i)^2} \leq 1 + \frac{4 \eta^\prime \eta_i+4\eta_o^2-4\eta_o(\eta^\prime+\eta_i)}{(\eta^\prime - \eta_i)^2}\,.
\end{align}
Combining this with relation~(\ref{large_delta}) in the form
$\exp(1/\epsilon)/\alpha\leq\delta^2$, we find the following strip for
the range of the integration variable $\eta^\prime$ that is allowed by causality
but where at the same time the approximation~(\ref{IR:approximation}) is invalid
due to a large separation between $\mathbf x^\prime$ and $\mathbf x_i$:
\beq \label{stripe}
\eta_i- 2 \sqrt{\alpha} {\rm e}^{- \frac{1}{2 \epsilon}}(\eta_o-\eta_i) \leq \eta^\prime \leq \eta_i + 2 \sqrt{\alpha} {\rm e}^{- \frac{1}{2 \epsilon}}(\eta_o-\eta_i)\,.
\eeq
We denote the restriction of the integral~(\ref{vertex_int})
to the strip~(\ref{stripe}) by $I_{\rm strip}$ and aim to determine an upper bound
on its absolute value.
Note that while relation~(\ref{stripe}) defines
an exponentially narrow strip around $\eta_i$, it becomes wider as $\eta_i$
takes large negative values. We therefore need to show in particular that
$|I_{\rm strip}|$ remains exponentially small even when $\eta_i\to -\infty$.
To estimate the integrand, we note that in the
region where the inequality~(\ref{large_delta}) holds, the term~(\ref{power_term})
appearing in the propagators satisfies the following relation:
\beq \label{up_bound_pow}
|y(x^\prime, x_i)|^{-\epsilon} \leq \alpha^{\epsilon} {\rm e}^{-1} \left( \frac{({\eta^\prime}^2 - \eta_i^2)}{\eta^\prime \eta_i} \right)^{- \epsilon}\,.
\eeq

\begin{figure}
  \centering
  \includegraphics[width=8cm]{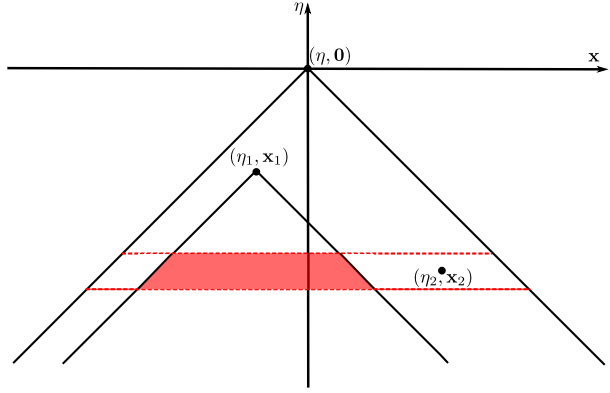}\\
  \caption{The vertex integration in Eq.~(\ref{vertex_int}) is confined to the past light cone of $x_1$. The large $\delta^2$ region for $y(x', x_2)$ is the narrow strip between the red dashed lines.}\label{fig:spacelike}
\end{figure}

Now, as illustrated in Figure~\ref{fig:spacelike},
the integration~(\ref{vertex_int}) over $x'$ is confined to the past light cone of $x_1$ because of the causal $\theta$ functions. This implies that the condition~(\ref{large_delta}) is never fulfilled for $\delta(x^\prime,x_1)$ and that we can
replace $\delta^2(x^\prime,x_1)$ with the approximation~(\ref{IR:approximation}).
In order to estimate contributions from the regions where the expansion~(\ref{IR:approximation}) breaks down,
we therefore consider the distance from $x^\prime$ to one of the remaining points, say $x_2$, where $\eta_2<\eta_1$.
Besides, we see from the relation~(\ref{stripe}) that the strip becomes wider and we obtain a looser
estimate on $I_{\rm strip}$ when taking $\eta_o\to0^-$, such that we set $\eta_o=0$ in order
to obtain an upper bound in the following.
In Figure~\ref{fig:spacelike}, the shaded area between the two dashed lines then indicates
the region where the inequality~(\ref{stripe}) is saturated because
of a large value of $\delta(x^\prime,x_2)$.
Assuming that $x_3$ and $x_4$ lie outside of the strip between the dashed lines, we can obtain an upper bound on $|I_{\rm strip}|$ by replacing the power term involving
$y(x^\prime,x_2)$ with the bound~(\ref{up_bound_pow}), while we make use of
Eq.~(\ref{power_term}) and the approximation~(\ref{IR:approximation}) for the
remaining power terms. We thus find:
\bea
|I_{\rm strip}| &\leq&  \frac{\lambda}{2H^2} \int_{\eta_2 - 2 \eta_2 \sqrt{\alpha} {\rm e}^{- \frac{1}{2 \epsilon}} }^{\eta_2 + 2 \eta_2 \sqrt{\alpha} {\rm e}^{- \frac{1}{2 \epsilon}}} d \eta' \frac{1}{\eta'^4} \int_0^{\eta_1 - \eta'} d r 4 \pi r^2 \alpha^\epsilon {\rm e}^{-1} \prod\limits_{i=1}^4
\left( \frac{(\eta_i - \eta')^2}{\eta_i \eta'} \right)^{- \epsilon}\nonumber \\
&\approx&  \frac{2\pi\lambda}{3H^2} {\rm e}^{-1} \frac{(\eta_1-\eta_2)^3}{\eta_2^4}
\prod\limits_{i=3,4}\left( \frac{(\eta_i - \eta_2)^2}{\eta_i \eta_2} \right)^{- \epsilon}
\int_{\eta_2 - 2 \eta_2 \sqrt{\alpha} {\rm e}^{- \frac{1}{2 \epsilon}} }^{\eta_2 + 2 \eta_2 \sqrt{\alpha} {\rm e}^{- \frac{1}{2 \epsilon}}} d \eta'
\left( \frac{(\eta_2 - \eta')^2}{\eta_2^2} \right)^{- \epsilon}
\nonumber \\
&\approx&
\frac{4\pi\lambda}{3H^2}\sqrt\alpha{\rm e}^{-\frac{1}{\epsilon}}
\frac{(\eta_1-\eta_2)^3}{\eta_2^3}
\prod\limits_{i=3,4}\left( \frac{(\eta_i - \eta_2)^2}{\eta_i \eta_2} \right)^{- \epsilon}\,,
\eea
which is exponentially small when compared with the estimate~(\ref{I:estimate}) for the leading IR
contribution.
Note in particular that this bound remains exponentially restrictive also for $\eta_2\to - \infty$.

Provided the individual $\eta_{2,3,4}$ are separated far enough, the above argument can be successively
applied as the $\eta^\prime$ integration sweeps over the disjoint strips~(\ref{stripe}).
We should eventually comment on situations where the individual strips defined by relation~(\ref{stripe}) overlap
or where these strips intersect with the light cones of the individual $x_i$. Since, as we have shown here,
the problematic regions
are exponentially small and the contained singularities integrable as long as $\epsilon \ll 1$, which
is amply fulfilled here by assumption, the contributions from intersections of
the regions~(\ref{vici_ligcon}) and~(\ref{large_delta}) are also exponentially small compared
to the leading IR terms.

In conclusion, the approximation (\ref{approximation}) can safely be used for the evaluation of Feynman diagrams to leading IR order and in the late-time limit. This implies in turn that in the same limit, the elements (propagators and vertices) of Feynman diagrams of the Amphichronous QFT are equivalent to their stochastic counterparts up to terms of order
$m^2/H^2$.

\section{Resumming the QFT in the Late-Time Limit}
\label{sec:resum}

In the previous Sections, we have demonstrated that the perturbative computations in QFT and in Starobinsky's stochastic approach agree at the leading IR order. In particular, there is a one-to-one correspondence between the Feynman diagram expansion of two-point functions
derived from the stochastic partition function~(\ref{def:go1}) and from Amphichronous
QFT in the Keldysh basis.
However, the stochastic approach offers the possibility to resum the perturbation series in terms of taking expectation values of a classical
probability distribution function ({\it i.e.} effectively in terms of a one-dimensional
integration), and consequently this may also serve as a resummation for all QFT diagrams
to leading IR order. In fact, the stochastic resummation yields well defined results at late times even in those cases when the perturbative expansion fails {\it i.e.}~for ultra-light and massless fields: $0\leq m^2\ll\sqrt \lambda H^2$. The result of this resummation was first obtained in the seminal work of Starobinsky and Yokoyama~\cite{Starobinsky:1994bd}. In the following,
we show that this resummation procedure can be applied as well
to the Feynman diagram expansions
presented in the present work. For this purpose, we demonstrate that in the late-time limit,
the stochastic partition function~(\ref{def:go1}) can effectively be evaluated with the same result as
the probability distribution found by Starobinsky and Yokoyama.

We start with the partition function $\mathcal{Z}$ as defined in Eq.~(\ref{def:go1}) and integrate
over the auxiliary field $\psi$. This yields

\beq
\mathcal{Z} = \int D[\phi] \, {\rm e}^{-\frac{2\pi^2}{H^3} \int_0^T \mathrm{d}u \, \left( \dot{\phi} +\frac{\partial_\phi V}{3H}\right)^2} = \int D[\phi] \, {\rm e}^{-\frac{2\pi^2}{H^3} \int_0^T \mathrm{d}t \, \left( \dot{\phi}^2 +2 \dot{\phi}\frac{\partial_\phi V}{3H} + \left(\frac{\partial_\phi V}{3H}\right)^2\right)}\,,
\label{eq:aux_compa_staro}
\eeq

\noindent where we have explicitly specified the boundaries of the integral in the exponential term. Note that in the above equation the potential is the full interacting one $V = \frac{1}{2}m^2\phi^2 + V_{\mathrm{int}}(\phi)$; we have also absorbed a constant in the integration measure. The partition function can then be written as

\beq\label{eq:partition-2}
\mathcal{Z} = \int_{-\infty}^{+\infty}\mathrm{d}\phi_T \,\, {\rm e}^{-\frac{4\pi^2}{3H^4}V(\phi_T)}\int D[\phi_{t<T}]{\rm e}^{-\frac{2\pi^2}{H^3} \mathcal{S}[\phi]}
\eeq
with
\beq
\mathcal{S}[\phi]=\int_0^T \mathrm{d}t\left[\dot{\phi}^2 + \left(\frac{\partial_\phi V}{3H}\right)^2\right]\,,
\label{def:action}
\eeq
%
\noindent where $\phi_T \equiv \phi(T)$, and we have assumed that $\phi(0) = 0$,  $V(0) = 0$.
Moreover, we have decomposed the measure in Eq.~(\ref{eq:partition-2}) as $\int D[\phi] = \int D[\phi_{t<T}]\int_{-\infty}^{+\infty}\mathrm{d}\phi_T$. The choice of initial condition implies that the IR sector of the field $\phi$ at $t=0$ does not exhibit significant fluctuations, {\it i.e.} there is no infrared enhancement yet. This could be, for example, due to inflation beginning at $t=0$. It is useful to note that $\mathcal{S}[\phi]$ corresponds to the action of a particle moving in the one dimensional potential $-[\partial_\phi V/(3H)]^2$ along the trajectory $\phi(t)$.

Our goal is to resum the QFT series and we are interested in the late time limit when this series will possibly break down. According to our previous arguments, the late-time limit should be given by evaluating the partition function (\ref{eq:partition-2}) non-perturbatively. This can be done by the steepest descent method. At late times, the integral is dominated by the path $\phi_0(t)$ that extremizes the pseudo action $\mathcal{S}$, {\it i.e.} $\left.\frac{\delta \mathcal{S}}{\delta\phi}\right|_{\phi = \phi_0}=0$. The latter condition is equivalent to

\beq
\ddot{\phi_0} - \left.\frac{1}{9H^2}\partial_\phi V \partial_\phi^2 V\right|_{\phi = \phi_0} = 0.
\eeq

\noindent Moreover the solution has to satisfy the following boundary
conditions: $\phi_0(0) = 0$ and $\phi_0(t) = \phi_t$. Integrating the above equation yields
\beq
\label{EOM:phi:integral}
\dot{\phi_0}^2 - \frac{1}{9H^2}\left(\left.\partial_\phi V\right|_{\phi = \phi_0}\right)^2 = E,
\eeq

\noindent where $E$ is an integration constant that should be chosen
in order to meet the boundary condition $\phi_0(T)=\phi_T$. For example, in the quartic potential
$V(\phi)=\frac{\lambda}{4!}\phi^4$, we have
$\partial_\phi V(\phi)|_{\phi=\phi_0(0)=0}=0$, and consequently, we have to take $E\to 0$ for $T\to \infty$,
corresponding to the fact that at $t=0$, we have to choose the kinetic energy of $\phi$ at the top of the
unbounded upside-down potential to be infinitesimally small for $\phi(T)$ to remain finite when
$T\to\infty$. More precisely, since $\phi(0)=0$, Eq.~(\ref{EOM:phi:integral}) implies that
$\phi(t)\geq \sqrt E t$. With the boundary condition $\phi(T)=\phi_T$, this implies that
$ET\leq \phi_T/T$ and therefore $ET\to0$ in the late-time limit $T\to \infty$.
%
\noindent
Therefore, in this limit, the pseudo action~(\ref{def:action}) can be written as
\begin{align}
\mathcal{S}[\phi_0] =& 2 \int_0^T \mathrm{d}t \, \dot{\phi_0}^2= \frac{2}{3H} \int_0^T \mathrm{d}t \, \dot{\phi_0}\left.\partial_\phi V\right|_{\phi = \phi_0}
= \frac{2}{3H} V(\phi_T)\,,
\label{eq:aux_compa_staro_3}
\end{align}

\noindent where we have used the boundary conditions $\phi_0(0) = 0$, $\phi_0(T) = \phi_T$,
and we have dropped terms of order $ET$. Using Eq.~(\ref{eq:aux_compa_staro_3}) in Eq.~(\ref{def:action}) and relabeling $\phi_T=\varphi$, we find
for the late time partition function
\beq\label{eq:partition-3}
\mathcal{Z} = \int_{-\infty}^{+\infty} \, \mathrm{d}\varphi \,\, {\rm e}^{-\frac{8\pi^2}{3H^4}V(\varphi)}\,.
\eeq

\noindent This coincides exactly with the result by Starobinsky and Yokoyama~\cite{Starobinsky:1994bd}. As mentioned above, the partition function (\ref{eq:partition-3}) is of course valid even for light
fields with $0\leq m^2\ll\sqrt \lambda H^2$, showing that their fluctuations remain well defined at late times.

Now one may note that the QFT calculations in this work assume $m>0$ through the
form of the free propagators~(\ref{KeldyshGF}). Since the stochastically
resummed result is continuous in the limit $m\to 0$ and on the other hand, the stochastic and the QFT diagrams agree, we may therefore conclude that the QFT result for the two-point
correlation coincides with the stochastic answer also for $m=0$. This can be also seen by
taking a different approach, {\it i.e.} by evaluating only diagrams that descend from the two-particle
irreducible effective action for the theory with $m=0$
and making a dynamical mass ansatz for the propagators. In Ref.~\cite{Beneke:2012kn},
it is demonstrated that these diagrams can be resummed to leading IR order in Euclidean de~Sitter space. This
leads to Schwinger-Dyson equations that can be solved for the dynamical mass, which is then found to be in agreement
with the result from Stochastic Inflation. Since at leading IR order, the QFT partition function in Euclidean de~Sitter space
coincides with the stochastic partition function, which we have shown here
in the same approximation also to be consistent
with
QFT in Lorentzian de~Sitter space, the resummation found in Ref.~\cite{Beneke:2012kn} can be directly applied to the
Feynman diagram expansions developed in the present work.

\section{Discussion} \label{sec:discussion}

In this paper we have demonstrated that for a light scalar field with quartic self-interaction, the Quantum Field Theory in
Lorentzian de~Sitter space and the stochastic theory of Starobinsky are in one-to-one correspondence at the level of Feynman diagrams in the IR limit, {\it i.e.} for spatial distances or time separations $\gsim1/H$. Corrections to this agreement will appear
at the relative order $m^2/H^2$ for a light massive scalar field ($\sqrt\lambda H^2\ll m^2\ll H^2$) and at order
$\sqrt{\lambda}$ for a massless or ultra-light field ($m^2\ll\sqrt\lambda H^2$).
Hence the stochastic formalism is perturbatively equivalent to the full QFT to leading IR order.
In addition to truncating
the Field Theoretical computations in a meaningful way, the stochastic approach offers the possibility of resumming the perturbation series in terms of late-time probability distribution functions.
This is of particular relevance for the ultralight or massless regime, where the perturbative
expansion breaks down. The equivalence at the level of Feynman diagrams implies that the stochastic resummation
can also be applied to QFT calculations to leading IR order~\cite{Starobinsky:1994bd}.

A number of comments on the relation between Field Theoretically and stochastically computed correlations
can be found in the earlier literature. It has indeed been conjectured and emphasized {\it e.g.}
in Refs.~\cite{Onemli:2002hr,Tsamis:2005hd} that the stochastic probability distribution function resums the
`leading logarithms' ({\it i.e.} the IR enhanced powers of $H^2/m^2$ in our nomenclature)
that occur in the Feynman diagrams of the Amphichronous Field Theoretical formulation.
Our analysis is the first
that fully reproduces the stochastic correlation functions found by Starobinsky and Yokoyama
from a QFT calculation truncated at leading IR order, and it therefore proves the
above conjecture.

While the main results of this paper as well as of many other
works concern equal-time correlations, it is also
interesting to consider their behaviour at unequal times. In the massive case,
the free theory
propagators~(\ref{free:propagators}) and~(\ref{KeldyshGF}) exhibit an
exponential decay on scales larger than $\sim 3H/m^2$. From the fact that
a dynamical mass ansatz for the Schwinger-Dyson equations derived
from the two-particle irreducible effective action in Euclidean de~Sitter space
reproduces the stochastic answers~\cite{Beneke:2012kn}, one may anticipate that this decay of correlations
at large separations should also appear in the massless theory on a Lorentzian metric. In particular,
when replacing $m^2\to m^2_{\rm dyn}=\Gamma(1/4)\sqrt\lambda H^2/(8\pi\Gamma(3/4))$, the relevant scale for the decay should be $3H/m_{\rm dyn}^2\sim H^{-1}/\sqrt\lambda$.
This may be compared with  Ref.~\cite{Kuehnel:2010}, where it is reported that
this scale should be $\sim H^{-1}(\lambda N)^{-1/3}$, where $N$ is the number of ${\rm e}$-folds. Moreover, it is found there that the decay of the correlations in momentum
space is rather abrupt toward large scales,
{\it i.e.} $\sim k^3$ where $k$ is the momentum. This should be
compared to a decay $\sim k^{m_{\rm dyn}^2/(3H^2)}$ indicated by the dynamical mass
ansatz, corresponding to a small and constant blue spectral tilt.
The answer for the decay of correlations in Lorentzian space
should be in principle
attainable from the stochastic functional~(\ref{def:go1}), a calculation which we leave
in detail to future work. Here, we note that a simple estimate appears to
support the decay behaviour suggested by the dynamical mass approach:
We take Eq.~(\ref{eq:starobinsky}) evaluated at the time $t$, multiply with $\phi(0)$ and take
the expectation value such that we get
\begin{align}
\langle\frac{d}{dt}\phi(t)\phi(0)\rangle+\lambda\langle\phi^3(t) \phi(0)\rangle/(18H)\approx 0\,,
\end{align}
where we assume that $t$ is large enough for the field and the noise to be uncorrelated,
$\langle\xi(t)\phi(0)\rangle\approx0$. Next, we Wick expand the correlation of
four fields to obtain
\begin{align}
\label{unequaltime:gauss}
\frac{d}{dt}\langle \phi(t)\phi(0)\rangle\sim \frac{\lambda}{H}\langle \phi^2(t)\rangle\langle\phi(t)\phi(0)\rangle\,.
\end{align}
Since the Wick expansion is valid only for approximately Gau{\ss}ian correlations,  above relation should be understood as an estimate
of order one accuracy for the
massless field in the quartic potential. When noting that the late-time limit of the
equal-time correlator follows from Eq.~(\ref{eq:partition-3}) to be~\cite{Starobinsky:1994bd}
\begin{align}
\langle \phi^2(t)\rangle=\frac{\Gamma\left(\frac74\right)H^2}{\pi\Gamma\left(\frac54\right)\sqrt\lambda}\,,
\end{align}
we find from relation~(\ref{unequaltime:gauss}) that
\begin{align}
\langle \phi(t)\phi(0)\rangle\sim {\rm e}^{- A \sqrt \lambda H t}\,.
\end{align}
The results from Ref.~\cite{Beneke:2012kn} for Euclidean de~Sitter space
suggest that
$A\sqrt \lambda H=m_{\rm dyn}^2/(3H)$. It would be interesting to explicitly
verify this conjecture from the stochastic partition function~(\ref{eq:aux_compa_staro})
that should in principle allow for an evaluation of unequal time correlators.

Future research may progress into directions on Euclidean de~Sitter space, {\it e.g.}
to aim for resummations of the massless scalar theory beyond the leading IR order or
for identifying the ground state of quantized Gravity, at least to leading IR approximation.
As for the developments on Lorentzian de~Sitter space performed in this work, these
open opportunities to address some of the following questions:
\begin{itemize}
\item
When going beyond the leading IR approximation, we expect also ultraviolet
divergences, that should be renormalized. One should investigate, whether the
known counterterms from the theory in Minkowski background are sufficient, or
if new operators that couple $\phi$ to scalar curvature invariants appear ({\it cf.}
Ref.~\cite{Arai:2012sh} for a discussion of such matters in the Hartree approximation).
This question is of importance for understanding the properties of scalar potentials
when receiving radiative corrections in the curved background.
\item
While the Starobinsky Equation~(\ref{eq:starobinsky}) in conjunction
with the stochastic noise~(\ref{eq:noise_correlation}) can be readily solved
in order to obtain the scalar field correlations at all times~\cite{Starobinsky:1994bd},
this has not yet been achieved in the QFT framework. In the perturbative
calculations~\cite{Onemli:2002hr,Onemli:2004mb,Brunier:2004sb,Kahya:2006hc}, early time correlations are addressed, while here, we show that
also the asymptotically time-independent correlations at late times can be obtained
using QFT methods, including situations where the perturbation expansion breaks down due
to the strong IR enhancement. It would be interesting to confirm the full evolution
of the correlations from early-time growth to late-time saturation for the
massless, self-interacting scalar theory within the QFT framework.
\item
Based on the conjecture that the stochastic approach sums all leading IR order diagrams,
it has been proposed to generalize its application beyond the self-interacting
scalar theory to {\it e.g.} Gravitation and Electrodynamics~\cite{Tsamis:2005hd,Prokopec:2007ak}.
It should be
interesting to use the methods developed here to establish the link between Field
Theoretical and stochastic formulations also for these well-motivated proposals.
\item
The stochastic probability distributions
in Ref.~\cite{Starobinsky:1994bd}
as well as the stochastic partition function~(\ref{def:go1}) make only predictions for the
time-dependence of the correlations. It would therefore be beneficial to
develop a formulation of the stochastic approach that can also make predictions
for the spatial dependencies. Ideally, this formulation should
be manifestly de~Sitter invariant, just as the underlying Field Theory.
\end{itemize}

While the stochastic approach  is
intuitive and compellingly simple, the Amphi\-chro\-nous QFT formulation
allows to perform systematic calculations
using controlled approximations.
To this end, the agreement found to leading IR order should be of practical use for validating stochastic
results by QFT calculations. In addition,
the links established here may be a starting point
to further develop calculational methods
for Quantum Theory on de~Sitter space that open paths to new results by
combining the advantages of both approaches.

\subsubsection*{Acknowledgements}
\noindent
The work of BG, FG and YZ is supported by the Gottfried Wilhelm Leibniz programme
of the Deutsche Forschungsgemeinschaft and by the DFG Cluster of Excellence Origin and Structure of the Universe. GR was supported by the Deutsche Forschungsgemeinschaft through the TRR33 program ``The Dark Universe''.

\bibliographystyle{heprint3}

\bibliography{feynqftstoch}

\end{document}